\documentclass[a4paper]{revtex4}
\usepackage{graphicx}
\usepackage{bbm}
\usepackage{amsmath,amssymb,amsfonts,amstext,latexsym,stmaryrd,amscd,txfonts,pxfonts}
\usepackage{ae}
\usepackage{bm}
\begin{document}

\title{On Quantum Cosmology in Teleparallel Gravity}

\author{A. S. Fernandes}
\address{Instituto de Física, Universidade de Brasília, 70910-900, Brasília, DF, Brazil}
\email{alexandre.fernandes@catolica.edu.br}

\author{S. C. Ulhoa}
\address{Instituto de Física, Universidade de Brasília, 70910-900, Brasília, DF, Brazil}
\email{sc.ulhoa@gmail.com}

\author{R. G. G. Amorim}
\address{Faculdade Gama, Universidade de Brasília, 72444-240, Setor Leste (Gama), Brasília, DF, Brazil.}
\email{ronniamorim@gmail.com}
\begin{abstract}
A quantum cosmology in teleparallel gravity is presented in this article. Teleparallel gravity is used to perform such an analysis once in General Relativity (GR) the concept of gravitational energy is misleading preventing the establishment of a concise quantum cosmology. The Wheeler-DeWitt like equation is obtained using the Weyl quantization and the teleparallel expression of energy.
\end{abstract}

\maketitle

\section{Introduction} \label{I}
The modern observational cosmology inaugurated at the Mount Wilson Observatory gave a great impetus to understanding the Universe \cite{Hubble.1929}. The Standard Cosmological Model, alongside the Cosmological Principle and field equations of GR,  describes all knowledge about large structures with good approximation. The Hubble Law shows how fast galaxies move away from each other at a relatively small distances. Thus it could be used to test new cosmological theories. The Cosmological Principle states that the Universe is isotropic (above 100 Mpc) and homogeneous (there is no center) in addition its dynamics is given by the Einstein field equations, $R_{\mu \nu} - \frac{1}{2} g_{\mu \nu} R = 8 \pi T_{\mu \nu}$. In such a way it is possible to trace a complete time evolution of the Universe. If the time is set backwards we see that everything started in a warm and dense state with domination of the radiation energy. The metric that admits the Cosmological Principle and the dynamics given by the GR is that of Friedman-Lemaître-Robertson-Walker (FLRW) \cite{Friedman.1999,Lemaitre.1927/31,Robertson.I,Robertson.II/III,Walker.1936}:
\begin{equation} \label{Eq-01}
	ds^2 = -dt^2 + a^2(t) \left[ \frac{dr^2}{1- \kappa r^2} + r^2 d \Omega^2\right], \, \qquad d \Omega ^2 \equiv d \theta + \sin ^2 \theta \, d \phi ^2
\end{equation}
where $k$ assumes values of $-1$ (negative or closed spatial curvature), $0$ (null or flat spatial curvature) or $+1$ (positive or open spatial curvature).

Due to the rapid improvement in observational cosmology a quantum theory of cosmology is increasingly needed in order to describe new effects. Particularly quantum effects should be relevant at singular states as it is expected in the beginning of the Universe. Such a theory should have an operator and time evolving structure. Since the seminal paper of Rosenfeld in 1930 \cite{Rosenfeld.1930,DeWitt.QTG.1970}, a lot of research has been done in this subject. It should be noticed that in order to proper treat a quantum theory of gravitation it is necessary to apply any quantization procedure to a physical quantity. On one hand the Hamiltonian is the most recognizable quantity with such a feature, on the other hand the existence of a gravitational energy is perhaps one of the most controversial and oldest problems in General Relativity. Thus a Hamiltonian formulation of GR was presented by Arnowitt, Deser and Misner in 1962 \cite {ADM.1962} and used by Wheeler and DeWitt  in 1967 to settle an Einstein-Schr\"{o}dinger equation \cite{DeWitt.QTG-I} which happened to be known as Wheeler-DeWitt equation. Such an equation has been focus of extensive research in the last 50 years. However the so called problem of time still persists \cite{Anderson.Time.WDW} which prevents a dynamical interpretation of one possible solution.

In this article we address the problem of the establishment of a quantum theory of gravitation applied to cosmology using teleparallel gravity and the Weyl quantization procedure. Such a procedure allows the mapping of any function into a operator structure. Thus it is not necessary to write the functions in phase space which is the standard procedure to quantize fields. The Weyl's prescription only require two parameters. Teleparallel gravity is an alternative theory of gravitation and entirely equivalent to General Relativity. Einstein was one of the first to use teleparallel gravity to investigate a form to unify gravity and electromagnetism \cite{Einstein.Teleparallel}. In 1961 M\o{}ller showed that it is impossible to cancel the gravitational field by coordinate transformation, this means that the pseudo-tensor approach should be avoided. Rescuing the initial idea of Einstein, M\o{}ller demonstrated that it was possible to obtain a gravitational energy density \cite{Moller.1961}. In 1963 Pellegrini and Plebanski found a Lagrangian formulation of the teleparallel gravity based on M\o{}ller \cite{Pellegrini.Plebanski}. Already in 1994 Maluf described the Hamiltonian formulation for Teleparallelism Equivalent to GR (TEGR) \cite{Maluf.1994}. Several applications of the TERG have been made as a description of the gravitational angular momentum \cite{Ulhoa.momento.angular}, neutron stars \cite{Ulhoa.Neutron.Stars}, quantum Schwarzschild spacetime \cite{Ulhoa.Amorim.2014}, gravitational waves \cite{Ulhoa.GW} among others.

This article is divided as follows: in section \ref{II} we briefly recall the ideas of teleparallel gravity using natural unities with $c=G=1$. We show how to obtain the energy-momentum tensor of gravitational and matter fields. In section \ref{III} we present the Weyl quantization procedure and apply it to the total energy of FLRW metric. As a consequence we establish two quantum cosmological equations. One quantizing the total energy density and the other using the total energy itself which is function only of the scale factor and its derivative. Finally in the last section we present our conclusions.

\section{Teleparallelism Equivalent to General Relativity} \label{II}
Teleparallel Gravity is dynamically equivalent to GR since the Lagrangian density of both theories differ only by a total divergence which means that the field equations are equivalent. Teleparallel gravity is constructed in terms of the tetrad field, $e^a \, _\mu$, which has two symmetries \cite{Maluf.1994,Maluf.2013.TERG,Aldrovandi.Pereira}. The latin indices ($a = (0), (i)$) represent the $SO(3,1)$ group while the greek indices stand for diffeomorphisms. Thus the tetrad field projects one symmetry into another. General Relativity is described in terms of the metric tensor that relates to tetrad field by $g_{\mu \nu} \, = \, e^a\,_{\mu} e_{a \nu}$.

Let's consider a geometry endowed with two connections, one ($\omega_{\mu a b}$) acting on Lorentz indices and another one ($\Gamma^\lambda\,_{\mu \nu}$) acting on transformations of coordinates. If we require that this space is metric compatible then $\nabla_\mu e_{a\nu}=0$. As a consequence $\Gamma^\lambda\,_{\mu \nu} = e^{a \lambda} e^b\,_\nu \omega_{\mu a b} + e^{a \lambda} \partial_\mu e_{a\nu}$. The curvature of this connection is $R^\lambda\, _{\gamma \mu \nu}(e,\omega) = e_a\,^\lambda e^b\,_\gamma (\partial_\mu \omega_{\nu a b} - \partial_\nu \omega_{\mu a b} + \omega_{\mu a c} \omega_{\nu c b}-\omega_{\nu a c} \omega_{\mu c b})$ and the torsion associated to such a connection is
\begin{equation} \label{Eq-02}
	T^\lambda\,_{\mu \nu} = \Gamma^\lambda \, _{\mu \nu} - \Gamma^\lambda \, _{\nu \mu}\,,
\end{equation}
which is explicitly given by
\begin{equation} \label{Eq-03}
	T^a\,_{\mu \nu} (e, \omega) = \partial_ \mu e^a\,_\nu - \partial_\nu e^a\,_\mu + \omega_{\mu a \nu} - \omega_{\nu a \mu} \, .
\end{equation}
The connection $\omega_{\mu a b}$ is known as the spin connection since it is used to couple gravity to a fermion field \cite{Maluf.2003.Dirac}. It is related to the Levi-Civita connection by means the mathematical identity
\begin{equation} \label{Eq-04}
	\omega_{\mu a b} =\, ^0 \omega_{\mu a b} + K_{\mu a b}
\end{equation}
where $^0 \omega_{\mu a b} = -\frac{1}{2} e^c\,_\mu (\Omega_{abc} - \Omega_{bac} - \Omega_{cab})$, with $\Omega_{abc} = e_{a \nu} (e_b\,^\mu \partial_\mu e_c\,^\nu - e_c\,^\mu \partial_\mu e_b\,^\nu)$, is the Levi-Civita connection which is a torsionless connection and  $K_{\mu a b} = -\frac{1}{2} e_a\,^\lambda e_b\,^\nu (T_{\lambda \mu \nu} + T_{\nu \lambda \mu} + T_{\mu \lambda \nu})$ is the contorsion tensor.

The so called teleparallel condition $\omega_{\mu a b} = 0$ establishes a geometric gravitational theory with a vanishing curvature and with the presence a torsion tensor. In other words the non-zero torsion is responsible for gravitational effects \cite{Maluf.2013.TERG}. The teleparallel condition allows one to relate the scalar curvature of the Levi-Civita connection to a scalar constructed out of the torsion tensor. Such a relations reads
\begin{equation} \label{Eq-05}
	e \, R(e) \equiv - \, e \, \left( \, \frac{1}{4} \, T^{abc} T_{abc} + \frac{1}{2} \, T^{abc} T_{bac} - T^a \, _a \, \right) + 2 \, \partial _\mu \left( \, e \, T^\mu \, \right)
\end{equation}
where $T^a \, _a = T^b \, _b \, ^a$ and
\begin{equation} \label{Eq-06}
	T^a\,_{\mu \nu}(e) = \partial_ \mu e^a\,_\nu - \partial_\nu e^a\,_\mu \, .
\end{equation}

Hence the teleparallel Lagrangian density is written as
\begin{equation}  \label{Eq-07}
	\mathcal{L}(e_{a \mu})= -\, k \, e \, \left( \, \frac{1}{4} \, T^{abc} T_{abc} + \frac{1}{2} \, T^{abc} T_{bac} - T^aT_a \, \right) - \mathcal{L}_M \, ,
\end{equation}
with $k = \frac{1}{16 \pi}$ and $\mathcal{L}_M$ is lagrangian density for matter fields. It should be noticed that the total divergence in (\ref{Eq-05}) was neglected since it gives no dynamical contributions. In addition one should take into account that the above Lagrangian density is equivalent to the Hilbert-Einstein Lagrangian density. As a consequence GR is dynamically equivalent to teleparallel gravity. Expression (\ref{Eq-07}) can be rewritten as $\mathcal{L}(e_{a \mu})= - \, k \, e \, \Sigma^{abc} T_{abc} - \mathcal{L}_M$ where $\Sigma^{abc}$ is defined as \cite{Maluf.1994}
\begin{equation} \label{Eq-08}
	\Sigma^{abc} = \frac{1}{4} \left( \, T^{abc} + T^{bac} - T^{cab} \, \right) + \frac{1}{2} \, \left( \, \eta^{ac}T^b - \eta^{ab}T^b \, \right) \, .
\end{equation}
The field equations are calculated from functional variation of the Lagrangian density $\mathcal{L}$ with respect to tetrads $e^{a \mu}$. They read
\begin{equation} \label{Eq-09}
	\partial _\lambda (e \, \Sigma^{a \mu \lambda}) - e \, \left( \, \Sigma^{b \lambda \mu} T_{b \lambda} \, ^a - \frac{1}{4} \, e^{a \mu} T_{bcd} \Sigma^{bcd} \right) = \frac{1}{4k} e \, T^{a \mu} \, .
\end{equation}
Due to the equivalence between teleparallel gravity and General Relativity the following important relationship is established
\begin{eqnarray} \label{Eq-10}
	\partial _\lambda (e \Sigma^{a \mu \lambda}) - e \, \left( \Sigma^{b \lambda \mu} T_{b \lambda} \, ^a - \frac{1}{4} \, e^{a \mu} T_{bcd} \Sigma^{bcd} \right) &=& \frac{1}{2} \, e \, \left[R^{a \mu}(e) - \frac{1}{2} \, e^{a \mu} R(e) \right] \, , \nonumber \\
	& & \nonumber \\
	R_{a \mu}(e) - \frac{1}{2} \, e_{a \mu} R(e) &=& \frac{1}{2 \, k} \, T_{a \mu} \, ,
\end{eqnarray}
where $T_{a \mu}$ is the energy-momentum tensor of matter fields. Therefore the Riemann geometry which is set in terms of a metric tensor and the curvature tensor is equivalent to the Weitzenb\"ock geometry which is established in terms of the tetrad field and the torsion in equation (\ref{Eq-06}). On the other hand given a metric tensor there are infinite possible tetrads.  This apparent paradox is solved by considering the physical interpretation of the tetrad field. Once the zero component of the tetrad field is always tangent along of a world line of an observer, it is interpreted as $e_{(0)}\,^\mu=u^\mu(\tau) = dx^\mu / d\tau$. Generalizing this idea the tetrad field can be established by the choice of the reference frame \cite{Maluf.2013.TERG,Maluf.2007.equiv}.

Equation (\ref{Eq-09}) can be expressed as
\begin{equation} \label{Eq-11}
	\partial_\nu \left( e \, \Sigma^{a \lambda \nu} \right) = \frac{1}{4} \, e \, e^a \, _\mu(t^\lambda\,_\mu + T^{\lambda \, \mu}) \, ,
\end{equation}
where $t^{\lambda \mu} = k \, (4 \, \Sigma^{bc\lambda} T_{bc} \, ^\mu - g^{\lambda \mu} \Sigma^{bcd} T_{bcd})$ is interpreted  as the momentum-energy tensor of the gravitational field. Since the tensor $\Sigma ^{a \lambda \nu}$ is antisymmetric in the last two indices, then $\partial _\mu \partial_\nu \left( e \Sigma^{a \lambda \nu} \right) \equiv 0$. Thereby,
\begin{equation} \label{Eq-12}
	\partial _\lambda \left[ e \, e^a \, _\mu (t^ \lambda \, _\mu + T^{\lambda \, \mu})\right] = 0 \, .
\end{equation}
In other words, this relation is a law of local conservation for the total energy-momentum tensor \cite{Maluf.2013.TERG}. This equation enables one to write the following continuing equation
\begin{equation} \label{Eq-13}
	\frac{d}{dt} \int_V d^3 \, x \, e \, e^a \, _\mu \, (t^{0 \mu} + T^{0 \mu}) = - \oint_S \, dS_j [e \, e^a \, _\mu (t^{j \mu} + T_{j \mu})] \, .
\end{equation}
As mentioned above the tensors $t^{\lambda \mu}$ and $T^{\lambda \mu}$ are gravitational momentum-energy tensor and matter momentum-energy tensor, respectively. Thus the total energy-momentum vector is
\begin{equation} \label{Eq-14}
	P^a = - \int_V d^3 \, x \, \partial_j \Pi^{aj} = - \oint_S d S_j \, \Pi^{aj} \, ,
\end{equation}
with
\begin{equation} \label{Eq-15}
	\Pi^{aj} = - \, 4 \, \kappa \, e \, \Sigma^{a0j} \, .
\end{equation}
This equation represents the total energy-momentum vector because $\Pi^{a\mu}$ is related to $t^{a\mu}$ and $T^{a\mu}$ by the field equations. It is interesting to notice that the above expression do represent the gravitational energy-momentum for vanishing matter fields. We also point out that the expression (\ref{Eq-14}) is invariant under coordinates transformations, but it is a vector under Lorentz transformation. Both features are mandatory rather than only desirable for any definition of energy-momentum vector. They are present in Special Relativity and there is no reason to abandoning them once dealing with a gravitational theory.

\section{Quantum cosmology in teleparallel gravity} \label{III}
Since 1925 the quantization procedures have long been known in the literature \cite{Ulhoa.Amorim.2014,Berezin,Twareque.Quant.Met,Giulini.Quant.Met}. In the 1920s another method of quantization was presented by Hermann Weyl \cite{Weyl.Book.1928}. From the mathematical point of view, such a procedure establishes a clear correspondence between classical functions and quantum operators. The Weyl quantization formalism will be used here in quantum cosmology. However before that let's briefly recall the most recognizable method in Quantum Mechanics.

\subsection{Canonical quantization} \label{III.A}
Canonical quantization is represented by a correspondence between the classical observables, which are real values of functions $f(p, \, q)$ (with $(p, \, q) = (p_i, \, q_i) \in \Re^n \times \Re^n$) in phase space to self-adjoint operators $Q_f$ in the Hilbert space $L^2(\Re^n)$. By the theorem of Stone and von Neumann \cite{Twareque.Quant.Met,vonNeumann.1955} the operators $H[f(q_j)] \psi \equiv \hat{q}_i \psi = q_j \psi \, $ and $ \, H[f(p_j)] \psi \equiv \hat{p}_i \psi = - \frac{i \, h}{2 \, \pi} \, \frac{\partial \psi}{\partial q_j} \, $ are the only ones acting in Hilbert space that satisfies the irreducibility condition and the commutation relationships $[Q_{p_j}, \, Q_{p_k}] =  [Q_{q_j}, \, Q_{q_k}] = 0 \,$ and $ \, [Q_{p_k}, \, Q_{p_j}] = \frac{ih}{2\pi} \delta_{jk}I$. A canonically classic system is composed by momentum and position coordinates, $q_i$ and $p_i$ respectively, with $i = 1, \, 2, \, \dots, \, n$ where $n$ is the number of degrees of freedom. The quantum states of system is evolved by Schr\"{o}dinger equation $i \, \hbar \, \frac{\partial \Psi (t)}{\partial t} = \hat{H} \Psi (t)$.

\subsection{Weyl quantization} \label{III.B}
Contrary to canonical quantization which requires the functions to be defined in the phase space, the Weyl's quantization allows the quantization of functions in curved spaces. In fact the method require at least two independent parameters and it is invertible and univocal. However it doesn't defines the quantum equation that the operators should obey.  It is maps a function $f$ of a classical system of $n$ variables denoted by $z_i$ into quantum operators $W[f(z_i)]$. Thus Weyl's map $W: f\rightarrow W[f(z_i)]$ is \cite{Ulhoa.Amorim.2014}
\begin{equation}  \label{Eq-16}
	W \, [f(z_i)] := \frac{1}{(2 \, \pi)^n} \int d^nk \, d^nz \, f(z_i) \, exp \, \left( i \, k^l \, (z_l - \hat{z}_l) \right) \, .
\end{equation}
The space formed by $W[z_i]\equiv\hat{z}_i$, with $i = 1, \, 2, \, \dots, \, n$ where $n$ is the number of degrees of freedom, is non-commutative. In others words, it exchanges local coordinates $z_i$ by hermitian operators $\hat{z}_i$ obeying the following commutation relation
\begin{equation} \label{Eq-17}
	[\hat{z}_i, \, \hat{z_j}] = i \, \alpha_{ij} \, ,
\end{equation}
where $\alpha_{ij}$ is a anti-symmetric quantity.

As an example this method, let's consider the function $f(q \, p) = qp$, in this specific case, we have $[\hat{q}, \, \hat{p}] = i \, \hbar$. If we apply equation (\ref{Eq-16}) then $f(q, \, p) \rightarrow W[f(q, \, p)]$ turns out
\begin{equation}  \nonumber
	W \, [f(q, \, p)] := \frac{1}{(2 \, \pi \, \hbar)^2} \int dk \,  d\eta \, f(q, \, p) \, exp \, \left\lbrace \frac{i}{\hbar} \, \left[ \, k \, (q - \hat{q}) + \eta \, (p-\hat{p}) \, \right] \, \right \rbrace  \, .
\end{equation}
Using the Baker-Campbell-Hausdorff formula \cite{Baker-Campbell-Hausdorff} it yields
\begin{equation}  \nonumber
	W \, [f(q, \, p)] = \frac{1}{(2 \, \pi \, \hbar)^2} \int dk  \, d\eta \, dq \, dp \, qp \, exp \, \left[ \, \frac{i}{\hbar} \, k \, (q - \hat{q}) \, \right] \, exp \, \left[ \, \frac{i}{\hbar} \, \eta \, (p - \hat{p}) \, \right] \, exp \, (- \, k\eta / 2) \, .
\end{equation}
The last equation, after simplification, becomes
\begin{equation}  \nonumber
	W \, [f(q, \, p)] = \frac{1}{(2 \, \pi \, \hbar)^2} \int dk  \, d\eta \, qp \, \left\lbrace \frac{1}{(2 \, \pi \, \hbar)} \int dk \, exp \left[  \, \frac{i}{\hbar} \, k \, (q - \hat{q} -\hbar / i \eta) \right]  \, \right\rbrace  \, exp \left( \, \frac{i}{\hbar} \, \eta \, (p-\hat{p}) \, \right) \, .
\end{equation}
Identifying delta function and using the property $\int f(x) \, \delta(x-a) \, dx = f(a)$, we obtain
\begin{eqnarray} \nonumber
	W \, [f(q, \, p)] &=& \frac{1}{(2 \, \pi \, \hbar)} \int dq  \, dp \, d\eta \, qp \, \delta \, (q - \hat{q} - \hbar \eta / 2 \, i) \, exp \, \left[ \frac{i}{\hbar} \, \eta \, (p - \hat{p}) \right] \nonumber \\
	& & \nonumber \\
	&=& \frac{1}{(2 \, \pi \, \hbar)} \int dp \, d\eta \, \hat{q}  p \, exp \, \left[ \frac{i}{\hbar} \, \eta \, (p - \hat{p}) \right] + (\hbar/ 2 \, i) \, \frac{1}{(2 \, \pi \, \hbar)} \, dp \, d\eta \, \eta \, p \, exp \left[ \, \frac{i}{\hbar} \, \eta \, (p - \hat{p}) \right] \nonumber \\
	& & \nonumber \\
	&=& \hat{q} \, \int dp \, p \, \delta \, (p - \hat{p}) + \frac{1}{(2 \, \pi \, \hbar)} \, \int dp \, d\eta \, (\hbar/ 2 \, i) \, \eta \, p \, exp \, \left[ \frac{i}{\hbar} \, \eta \, (p - \hat{p}) \right] \, . \nonumber
\end{eqnarray}

Using the property $\int f(x) \, \frac{d \delta(x - a)}{d \, x} \, dx = - f'(a)$, and identifying the delta function again, it yields
\begin{equation}\nonumber
	W \, [f(q, \, p)] = \frac{1}{2}(\hat{q} \, \hat{p} + \hat{p} \, \hat{q}) \, ,
\end{equation}
as expected.

\subsection{Quantum cosmology} \label{III.C}
In order to construct a quantum cosmology we will calculate the energy of FLRW metric in (\ref{Eq-01}) and, using the teleparallel expression (\ref{Eq-15}), we'll quantize it by means the Weyl's method. The tetrad adapted to a stationary reference frame that yields the FLRW metric is given by
\begin{equation} \label{Eq-18}
	e ^{a} \, _ \mu= \begin{pmatrix}	
		1 & 0 & 0 & 0 \\
		0  & \frac{a}{\sqrt{1-kr^2}} \, sin\theta \, cos\phi & a \, r \, cos\theta \, cos\phi & - \, a\, r \, sin\theta \, \sin\phi \\
		0  & \frac{a}{\sqrt{1-kr^2}} \, sin\theta \, sin\phi & a \, r \, cos\theta \, sin\phi &  a \, r \, sin\theta \, cos\phi \\
		0  & \frac{a}{\sqrt{1-kr^2}} \, cos\theta & -a \, r sin\phi & 0
	\end{pmatrix} \, .
\end{equation}
The relevant component of $\Sigma^{abc}$ to calculate the energy is \cite{Maluf.2013.TERG}
\begin{equation} \label{Eq-19}
	\Sigma^{001} = \frac{1}{4} \, (2g^{11} \, T_{001}) - \frac{1}{2} \, g^{11} \, (T_{001} + g^{22} \, T_{212} + g^{33} \, T_{313}) \, ,
\end{equation}
where we used the relation $T^\beta = T^\alpha \, _\alpha \, ^\beta = g^{\alpha \lambda} g^{\beta \gamma} T_{\lambda \alpha \gamma}$.
Then after some algebraic manipulations we arrive at
\begin{equation} \label{Eq-20}
	\Sigma^{001} = \frac{\sqrt{1 - k \, r^2} - (1 - k \, r^2)}{a^2 \, r} \, .
\end{equation}
The total energy density of the Universe is
\begin{equation} \label{Eq-21}
	\Pi^{(0)1} = 4 \, \kappa \, a \, r \, \sin \theta \, \left(\sqrt{1 - k \, r^2} - 1 \right)
\end{equation}
where $\kappa \equiv \frac{1}{16 \pi}$, $k$ spatial curvature and $a = a(t)$ is the scale factor.

Let's firstly quantize the energy density, thus $\Pi^{(0)1}=\epsilon(r, \, \tau) = A \, r \, \tau \left( \sqrt{1 - k \, r^2} \, - \, 1 \right)$ where $A = 4 \, a \, \kappa$ and $\tau = \sin \theta$. Then the Weyl procedure yields $\hat{\mathcal{H}} = W[\epsilon(r, \, \tau)]$, it reads
\begin{equation} \label{Eq-22}
	\hat{\mathcal{H}} = \left[\frac{A}{2} \, (\hat{r} \, \hat{\tau} - \hat{\tau} \, \hat{r}) \right] \left(- \, k \, \hat{r}^2 + 1 \right)^{\frac{1}{2}} - \left[ \frac{A}{2} \, (\hat{r} \, \hat{\tau} - \hat{\tau} \, \hat{r}) \right]\,.
\end{equation}
If we use $\hat{r} = r$ and $\hat{\tau} =-i\alpha\frac{\partial}{\partial r}$, then $[\hat{r}, \, \hat{\tau}]=i\alpha$. This implies that the Hamiltonian density can be expanded in terms of the non-commutative parameter $\alpha$ which is supposed to be very small. As a consequence we overcome the problem of dealing with a squared operator. Hence
\begin{equation} \label{Eq-23}
	\hat{\mathcal{H}} = 3 \, a \, \kappa \, k \, i \, \alpha \, r^2 + 2 \, a \, \kappa \, k \, r^3 \, i \, \alpha \, \frac{\partial}{\partial r}\,.
\end{equation}
We assume that the dynamics is given by a Schr\"odinger like equation $\hat{\mathcal{H}\Psi}=i\hbar\frac{\partial\Psi}{\partial t}$, therefore
\begin{equation}\label{Eq-24}
	3 \, a \, \kappa \, k \, \alpha \, r^2 \, \Psi + 2 \, a \, \kappa \, k \, r^3 \, \alpha \, \frac{\partial \Psi}{\partial r} = \hbar \, \frac{\partial\Psi}{\partial t} \, .
\end{equation}
It should be noted that in the above equation the scale factor acts as a parameter. It is possible to solve the spatial equation which reads $$3 \, \kappa \, k \, \alpha \, r^2 \, \psi + 2 \, \kappa \, k \, r^3 \, \alpha \, \frac{\partial \psi}{\partial r} = \xi \psi \, ,$$ but the temporal part, $$\hbar \, \frac{\partial\phi}{\partial t} = a \, \xi \phi\,,$$ requires an explicit form of the scale factor. Here $\xi$ is the constant of separation of variables $\Psi(r,t) = \psi(r) \, \phi(t)$. Possibly the use of an altered temporal scale can solve the problem. Another attempt would be using a semi-classical approach in which the scale factor comes up from the Einstein equation.

The next possibility is to quantize the energy $P^{(0)} = E$ itself. Integrating the energy density over a spherical 3D surface with radius $r_0$ we get
\begin{equation} \label{Eq-25}
	E = -\lim\limits_{r \rightarrow r_0} \, 8 \, \pi \, \kappa \, a \, r \left[ \sqrt{1 - k \, r^2} \, - \, 1 \right]\,.
\end{equation}
It should be noted that $r_0$ is arbitrary. It specifies the region where we want to calculate the energy. Thus we can choose it to be the radius of the dynamical horizon of FLRW metric, $r_0 = \frac{1}{\sqrt{H^2 \, + \, \frac{k}{a^2}}}$ with $H = \frac{\dot{a}}{a}$. In this way $E$ will represent the energy of the observable Universe. In order to obtain an expression that describes the beginning of the Universe, we expand the energy as
\begin{equation} \label{Eq-26}
	E \, (a, \, \dot{a}) \approx \frac{2 \, a \, k \, \sqrt{k} \, - \, 2 \, a^2 \, k \, \sqrt{k} \, - \, a^3 \, k \, \sqrt{k} \, + \, \frac{1}{2} \, a^4 \, (\dot{a})^2 \, \sqrt{k} \, + \, a^6 \, k \, \sqrt{k}}{2 \, k^2} \, .
\end{equation}
Using the Weyl's method the Hamiltonian $\hat{H} = W \, [E]$ reads
\begin{equation} \label{Eq-27}
	\small{ 2 \, k^2 \, \hat{H} \approx 2 \, \hat{a} \, k \, \sqrt{k} \, - \, 2 \, \hat{a}^2 \, k \, \sqrt{k} \, - \, \hat{a}^3 \, k \, \sqrt{k} \, + \, \frac{1}{2} \, \hat{a}^4 \, \sqrt{k} \, (\hat{\dot{a}})^2 \, + \, \hat{a}^6 \, k \, \sqrt{k} \, - \, 3 \, \omega^2 \, \hat{a}^2 \, \sqrt{k} \, - \, i \, 2 \, \omega \, \hat{a} \, \sqrt{k} \, \hat{\dot{a}} } \, ,
\end{equation}
where $[\hat{a}, \, \hat{\dot{a}}] = i \, \omega$ with $\hat{a} = a$, $\hat{\dot{a}} = - \, i \, \omega\frac{\partial}{\partial a}$. Therefore following the Wheeler-DeWitt prescription $\mathcal{H} \Psi = 0$ we get
\begin{equation} \label{Eq-28}
	\small{ \left( 2 \, a \, k \, \sqrt{k} \, - \, 2 \, a^2 \, k \, \sqrt{k} \, - \, a^3 \, k \, \sqrt{k} \, - \, \frac{1}{2} \, \omega^2 \, a^4 \, \sqrt{k} \, \frac{\partial^2}{\partial a^2} \, + \, a^6 \, k \, \sqrt{k} \, - \, 3 \, \omega^2 \, a^2 \, \sqrt{k} \, - \, 2 \, \omega^2 \, a \, \sqrt{k} \, \frac{\partial}{\partial a} \right) \Psi = 0 } \, .
\end{equation}
Such an equation has the same dependence of the WDW equation and has a regime of validity for a small Universe.

\section{Conclusion} \label{IV}
In this article we have derived two quantum equations for cosmology. One of them was obtained by quantizing the energy density and the another one quantizing the energy itself. The quantization of energy leads to a Wheeler-DeWitt like equation, which is valid for a small Universe. However a similar equation should hold for the Universe's present time since the non-commutative parameter should be very small.
Since the publication of the RG the vision about the world has changed. With the coming of Friedman \cite{Friedman.1999}, Lemaître \cite{Lemaitre.1927/31}, Robertson \cite{Robertson.I,Robertson.II/III}, Walker \cite{Walker.1936} and Hubble \cite{Hubble.1929} the cosmological knowledge was raised to a very high plateau. With the coming of Quantum Physics the picture of the Universe is almost complete. What was missing? A union between GR and Quantum Physics.  However we believe that GR is not suitable theory to achieve such an aim, since it is not enough to deal with the problem of gravitational energy. On the other hand in the framework of teleparallel gravity it is possible to obtain a well defined expression for energy. Using the Weyl quantization, which is a procedure with a natural correspondence between the classical and quantum operators, the quantum gravity is established \cite{Ulhoa.Amorim.2014} and here applied to an Universe governed by the cosmological principle.


\end{document}